\documentclass[aps,prl,twocolumn,groupedaddress,showpacs]{revtex4}

\usepackage{amsmath}
\usepackage{amssymb}
\usepackage{amsfonts}
\usepackage{graphicx}

\newcommand{\absq}{\vert q^2\vert}
\newcommand{\cl}{\mathrm{cl}}
\newcommand{\qu}{\mathrm{qu}}
\newcommand{\hr}{\mathrm{harm}}
\newcommand{\Eq}{Eq.\,}
\newcommand{\Fig}{Fig.\,}
\newcommand{\Ref}{Ref.\,}

\begin{document}

\title{Quantum corrections to the Schwarzschild metric
and reparametrization transformations}

\author{G.G. Kirilin}
\email[]{G.G.Kirilin@inp.nsk.su}

\affiliation{Budker Institute of Nuclear Physics SB RAS, Novosibirsk, Russia}

\date{\today}

\begin{abstract}
Quantum corrections to the Schwarzschild metric generated by loop diagrams have
been considered by Bjerrum-Bohr, Donoghue, and Holstein (BHD) [Phys. Rev.
\textbf{D68}, 084005 (2003)], and Khriplovich and Kirilin (KK) [J. Exp. Theor.
Phys. \textbf{98}, 1063 (2004)]. Though the same field variables in a covariant
gauge are used, the results obtained differ from one another. The reason is
that the different sets of diagrams have been used. Here we will argue that the
quantum corrections to metric must be independent of the choice of field
variables, i.\,e. must be \textit{reparametrization invariant}. Using simple
reparametrization transformation, we will show that the contribution considered
by BDH, is not invariant under it. Meanwhile the contribution of the complete
set of the diagrams, considered by KK, satisfies the requirement of the
invariance.
\end{abstract}

\pacs{04.60.-m}

\maketitle

\section{General structure of reparametrization transformation}

In the series of papers Weinberg
\cite{Weinberg1964,Weinberg1964a,Weinberg1965}, Boulware and Deser
\cite{Boulware1975} have shown that the massless particles of helicity $\pm 2$
are described by the effective theory, satisfying the equivalence principle.
Boulware and Deser have shown that the corresponding effective action coincides
with the classical Einstein action~\footnote{We put $\hbar=c=16\pi G=1$,
restoring the dimension in the final results only.}
\begin{align}
\bar{S}_g=\int d^4x \left[-\sqrt{-\bar{g}}\,\bar{R}(\bar{g})\right].\label{e1}
\end{align}
In general case, it is necessary to supplement the lagrangian density with a
number (may be infinite) of terms of higher orders in $\partial_\alpha
\bar{g}_{\mu\nu}$. The natural property of any effective theory is the
\textit{reparametrization invariance}. It implies that a scattering amplitude
on mass shell does not depend on the choice of field variables. In general
relativity one of natural parametrizations of the gravitational field
$h_{\alpha\beta}$ is the decomposition of the covariant metric tensor:
$\bar{g}_{\mu\nu}=g_{\mu\nu}+f_{\mu\nu}(h)$, where $f$ is an arbitrary
symmetric tensor function, the expansion of which begins with a linear in
$h_{\alpha\beta}$ term. For example, to derive the counter lagrangian of the
gravity, interacting with a massless scalar field, 't\,Hooft and Veltman
\cite{Hooft1974} have used the trivial parametrization
\begin{align}
\bar{g}_{\mu\nu}=g_{\mu\nu}+h_{\mu\nu},\label{e2}
\end{align}
where $g_{\mu\nu}$ is the background field, $h_{\mu\nu}$ is the operator field,
characterizing quantum fluctuations. The action of scalar field in external
gravitational field has the form
\begin{align}
\bar{S}_{m}=\int d^{4}x\,\frac{\sqrt{-\bar{g}}}{2}\left(  \bar
{g}^{mn}\partial_{n}\bar{\phi}\,\partial_{m}\bar{\phi}-m^{2}\bar{\phi}%
^{2}\right).\label{e3}
\end{align}
Similarly to (\ref{e2}), we decompose the field $\bar{\phi}$
\begin{align}
\bar{\phi}=\tilde{\phi}+\phi.\label{e4}
\end{align}
Supplementing the action (\ref{e1}) with a gauge fixing part
\begin{align}
S_{f}=\int d^{4}x\,\frac{\sqrt{-g}}{2} \left(  h_{\mu
\mid\alpha}^{\alpha}{}-\frac{1}{2}h_{\mid\mu}\right)  \left(h_{\mid\beta
}^{\mu\beta}{}-\frac{1}{2}h^{\mid\mu}\right),\label{e5}
\end{align}
and a corresponding action of ghosts $\eta_\mu$, we find the expansion of the
action up to second order in fluctuations:
\begin{align}
\bar{S}_{g}+\bar{S}_{m}+S_{f}+S_{gh}(\eta)=\int d^{4}x\sqrt{-g}\left(
\mathcal{L}+\underline{\mathcal{L}}+\underline
{\underline{\mathcal{L}}}\right),\label{e6}
\end{align}
with
\begin{align}
\mathcal{L}  & = - R+\frac{1}{2}\left(  g^{\mu\nu}\partial_{\nu}\tilde{\phi}%
\,\partial_{\mu}\tilde{\phi}-m^{2}\tilde{\phi}^{2}\right),  \label{e7}\\
\underline{\mathcal{L}}  &
=\left(R_{\mu}^{\nu}-\frac{1}{2}\delta_{\mu}^{\nu}R-\frac
{1}{2}T_{\mu}^{\nu}\right)h^{\mu}_{\nu}
+\left(\tilde{\phi}_{\mid\lambda}^{\mid\lambda}+m^{2}\tilde{\phi}\right)
\phi,\label{e8}\\
\underline{\underline{\mathcal{L}}}  & =-\frac{1}{4}\left(  h_{\alpha\beta
}P^{\alpha\beta}_{\gamma\delta}{}h^{\gamma\delta}{}_{\mid\lambda}^{\mid\lambda
}+h_{\alpha\beta}X^{\alpha\beta}_{\gamma\delta}h^{\gamma\delta}\right),\label{e9}\\
& +\frac{1}{4}h_{\alpha\beta}W^{\alpha\beta}_{\gamma\delta}h^{\gamma\delta}
+\eta^{\dagger\,\mu}\left(  \eta_{\mu}{}^{\mid \lambda}{}_{\mid \lambda}+R_{\mu}^{\lambda}%
\eta_{\lambda}\right)\notag\notag,\\
& +\phi\left(  P^{\mu\nu}_{\gamma\delta}\partial_{\mu}\tilde{\phi}\,D_{\nu
}+P^{\mu\nu}_{\gamma\delta}\tilde{\phi}_{\mid\mu\nu}-\frac{1}{2}\,g_{\gamma
\delta}m^{2}\tilde{\phi}\right) h^{\gamma\delta}.\notag
\end{align}
In expressions (\ref{e7})-(\ref{e9}) indices are raised and lowered by means of
the tensor $g_{\mu\nu}$, $R_{\mu\nu}$ and $R$ are the Ricci tensor and Riemann
curvature of the background field, respectively. We introduce also the notation
$h=h^\lambda_{\lambda}$. Indices following vertical lines denote covariant
derivatives relative to the metric tensor $g_{\mu\nu}$. Matrices, appearing in
the expressions (\ref{e7})-(\ref{e9}), have the following form:
\begin{align}
P_{\gamma\delta}^{\alpha\beta}  & =\delta_{(\gamma}^{\alpha}\delta_{\delta
)}^{\beta}-\frac{1}{2}g^{\alpha\beta}g_{\gamma\delta}\,,\label{e9a}
\\
X^{\alpha\beta}_{\gamma\delta}  & =P^{\alpha\beta}_{\rho\sigma}\left[
R_{(\gamma }{}^{\rho}{}_{\delta)}{}^{\sigma}+\delta^{\sigma}_{(\delta}\left(
R^{\rho
}_{\gamma)}-\frac{1}{2}\delta^{\rho}_{\gamma)}R\right)  \right]  \notag\\
& +(\alpha\beta \leftrightarrow\gamma\delta)\,,\label{e10}\\
W^{\alpha\beta}_{\gamma\delta}  & = T^{(\alpha}_{\sigma}\,
P^{\sigma\beta)}_{\gamma\delta}+T_{(\gamma}^{\sigma}\,
P_{\sigma\delta)}^{\alpha\beta}\notag\\
& +\frac{1}{2}P^{\alpha\beta}_{\gamma\delta}\left(  m^{2}\tilde{\phi}%
^{2}-T\right).\label{e11}
\end{align}
In the expressions (\ref{e9a})-(\ref{e11}) indices with brackets are to be
symmetrized. $T_{\mu\nu}$ is the stress tensor of the scalar field:
\begin{align}
T_{\mu\nu}=\partial_{\mu}\tilde{\phi}\,\partial_{\nu}\tilde{\phi}-\,\frac{1}%
{2}\,g_{\mu\nu}\left(  g^{\rho\sigma}\partial_{\rho}\tilde{\phi}\,\partial
_{\sigma}\tilde{\phi}-m^{2}\tilde{\phi}^{2}\right).
\end{align}
The first variation of the action (\ref{e8}) eventually supply us with the
equations of motion for the background fields:
\begin{align}
& R_{\mu\nu}-\frac{1}{2}\,g_{\mu\nu}R=\frac
{1}{2}\,T_{\mu\nu}\,,\label{e12}\\
&g^{\mu\nu}\tilde{\phi}_{\mid\mu\nu}+m^{2}\tilde{\phi}=0\,.\label{e13}
\end{align}
In \Ref\cite{Boulware1975} it has been shown that, at fixed gauge, the three
graviton vertex is matched by the gravitational interaction with stress tensor
of the classical free spin 2 field up to four parameters, corresponding to the
reparametrization of the field $h_{\mu\nu}$~\footnote{There are linear changes
of variables such as $h_{\mu\nu}\to c_1 h_{\mu\nu}+c_2 g_{\mu\nu} h$, but we
leave them aside for the sake of simplicity.}:
\begin{align}
& \bar{g}_{\mu\nu}=g_{\mu\nu}+h_{\mu\nu}+a_1\,h_{\mu\lambda}h^{\lambda}_{\nu}\notag\\
& + a_2\,h_{\mu\nu} h + a_3\,g_{\mu\nu} h^{\alpha}_{\beta} h_{\alpha}^{\beta}+
a_4\,g_{\mu\nu} h^2. \label{e14}
\end{align}

\begin{figure}
\includegraphics{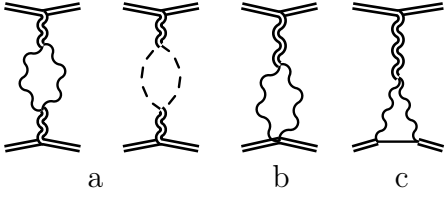}
\caption{The diagrams taken into account in \Ref\cite{Bjerrum-Bohr:2002ks}
\label{f1}}
\end{figure}

Loop corrections to the scattering amplitude have been studied in
Refs.\,\cite{Bjerrum-Bohr:2002ks}, \cite{Khriplovich:2004cx}. Corrections
concerned were proportional to $\ln\absq$, where $q^2$ is the transfer momentum
squared. In particular, it was found that, after averaging over the
fluctuations, corrections to the Schwarzschild metric appeared:
\begin{align}
g_{\mu\nu}=g^{\cl}_{\mu\nu}+\underline{g}{}_{\mu\nu}\,,\label{e15}
\end{align}
where $g^{\cl}_{\mu\nu}$ is the classical Schwarzschild solution,
$\underline{g}{}_{\mu\nu}$ is the quantum correction to it.
Quite apparently, the \textit{leading} corrections to metric must be
independent of the way of parametrization of the field $h_{\mu\nu}$. Actually,
being quadratic in fluctuations, additional terms in the parametrization
(\ref{e14}) generate additional structures to
$\underline{\underline{\mathcal{L}}}$ only due to replacement of the field
$h_{\mu\nu}$ in the lagrangian density $\underline{\mathcal{L}}$, 
i.\,e., these structures vanish after taking into account the equations of
motion (\ref{e12}). However, in perturbation theory it happens only if
\textit{all diagrams} have been taken into account. In
\Ref\cite{Bjerrum-Bohr:2002ks} only certain part of the diagrams have been
considered, namely, the graviton propagator corrections and the corrections to
one of the vertices (\Fig\ref{f1}). As we will show, the contribution of these
diagrams is not reparametrization invariant.

\section{Example of reparametrization transformation}

As an example, we parametrize the gravitational field in the following way
\begin{align}
\bar{g}_{\mu\nu}=g_{\mu\nu}+h_{\mu\nu}-\frac{a}{4}\,h_{\mu\alpha}h^\alpha_\nu.\label{e16}
\end{align}
As stated above, the lagrangian quadratic in fluctuation changes due to the
linear terms only. The reparametrization (\ref{e16}) is equivalent to the
replacement of the matrices $X$ and $W$ in the lagrangian density (\ref{e9}) by
the matrices $X+a \mathcal{X}$ and $W+a \mathcal{W}$, respectively, there
\begin{align}
\mathcal{X}_{\gamma\delta}^{\alpha\beta}  & =\delta_{(\gamma}^{(\alpha
}P^{\beta)}{}_{\delta),\kappa\lambda}R^{\kappa\lambda}\,,\label{e17}\\
\mathcal{W}_{\gamma\delta}^{\alpha\beta}  & =\frac{1}{2}\delta_{(\gamma
}^{(\alpha}T_{\delta)}^{\beta)}\,.\label{e17a}%
\end{align}
Graviton propagator corrections are generated by the counter lagrangian of pure
gravity. The counter lagrangian has been derived in~\Ref\cite{Hooft1974}, we
aim here to find its transformation under the reparametrization transformation
(\ref{e16}). Using the general formula for the counter lagrangian derived
in~\Ref\cite{Hooft1974}, we find:
\begin{align}
L_{count.}^{(a)} & =\frac{\sqrt{-g}}{8\pi^{2}(d-4)}\frac{1}%
{4}\,Sp\left\{2a\left(  X\mathcal{X}\right)\right.\notag\\
& \left. +\frac{a}{3}R\left( P\mathcal{X}\right)
+a^{2}\left(P\mathcal{X}P\mathcal{X}\right)\right\}\,.\label{e18}
\end{align}
In the expression (\ref{e18}) the matrices $P, X$ and $\mathcal{X}$ should be
read as $10\times 10$ matrices in relation to the number of the components of
the symmetric tensor $h_{\mu\nu}$. Adding up the results
of~\Ref\cite{Hooft1974} and (\ref{e18}) yields the counter lagrangian for the
case of pure gravity
\begin{align}
L_{count.}  & =\frac{\sqrt{-g}}{8\pi^{2}(d-4)}\left[  \left(
\frac{7}{20}+\frac{a^{2}}{8}\right)  R_{mn}R^{mn}\right.  \\
& \left.  +\left(  \frac{1}{120}+\frac{a}{8}\left(  \frac{14}{3}%
+ a\right)  \right)  R^{2}\right]  \,.\label{e18a}
\end{align}
This lagrangian gives the following corrections to the pure time component of
the metric (diagram \Fig\ref{f1}a):
\begin{align}
\underline{g}{}^{\ref{f1}a}_{00}= - \left[\frac{43}{15}+a\left(
\frac{14}{3}+2a\right) \right]  \frac{G^{2}\hbar m}{\pi
c^{5}r^{3}}\,.\label{e19}
\end{align}
Using the additional vertices (\ref{e17}), (\ref{e17a}), it is easy to find the
contributions of the diagrams depicted in Figs.\,\ref{f1}b,\,c:
\begin{align}
\underline{g}{}^{\ref{f1}b}_{00}  & =\left[  \frac{26}{3}+a\left(
\frac{37}{3}+2a\right)
\right]  \frac{G^{2}\hbar m}{\pi c^{5}r^{3}}\,,\label{e20}\\
\underline{g}{}^{\ref{f1}c}_{00}  & =-\left(  \frac{5}{3}+5a\right)
\frac{G^{2}\hbar m}{\pi
c^{5}r^{3}}\,.\label{e21}%
\end{align}
Summing up the results (\ref{e19})-(\ref{e21}), we get the following
contributions of the diagrams \Fig\ref{f1}:
\begin{align}
\underline{g}^{\ref{f1}a+\ref{f1}b+\ref{f1}c}_{00}=\left(
\frac{62}{15}-\frac{2}{3}\,a\right) \frac {G^{2}\hbar m}{\pi c^{5}r^{3}}\,.
\label{e22}
\end{align}
The a-independent part of \Eq(\ref{e22}) coincides with the result of
\Ref\cite{Bjerrum-Bohr:2002ks}. From \Eq(\ref{e22}) one can see that this
contribution is \textit{not reparametrization invariant}. Whereas the sum of
the contributions of all the diagrams, listed in \Ref\cite{Khriplovich:2004cx},
is {reparametrization invariant} for the obvious reason stated above
\begin{align}
\underline{g}^\qu_{00}=\frac{107}{30}\frac {G^{2}\hbar m}{\pi
c^{5}r^{3}}\,.\label{e23}
\end{align}
Parametrization dependence on the contribution of the diagrams \Fig\ref{f1}
(i.\,e., diagrams containing a single graviton propagator attached to one of
the particles) is the direct consequence of the fact that, in general
relativity, separation of these diagrams from other loop ones is a matter of
convention only, because \textit{they do not contain the pole} in $q^2$
\footnote{In contrast to QED or QCD these corrections lead to the
renormalization of the operators of higher dimensions than (\ref{e1}), for
example (\ref{e18a}), thus they bear no relation to the renormalization of
$G$.}. Being unrelated to the renormalization of the amplitude with pole in
$q^2$, these diagrams should be considered in line with other ones. As it has
been shown, reparametrization transformations mix this diagrams with, for
example,  diagram proportional to $Sp\{PWPW\}$ (see Eqs.\,(\ref{e9a}),
(\ref{e11})). Due to \Eq(\ref{e12}) there is no difference between the
contribution of the diagrams \Fig\ref{f1} on mass shell and, for example, the
diagram proportional to $Sp\{PWPW\}$.

\section{Aside on classical corrections}

\begin{figure}
\includegraphics{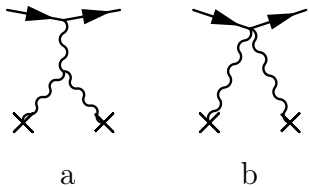}%
\caption{Tree diagrams \label{f2}}
\end{figure}

The correction (\ref{e23}) is the leading one in $l_p^2/r^2$, where $l_p$ is
the Plank length. From the standpoint of leading corrections, the
parametrizations (\ref{e2}), (\ref{e16}) are indeed indistinguishable, because,
after averaging over the quantum fluctuation, the information about
parametrization of these fluctuation is lost.
Therein lies the main difference between the leading quantum corrections and
nonleading classical corrections of the order $r^2_g/r^2$, there $r_g$ is the
Schwarzschild radius. Let us consider this aspect in detail. The diagram
depicted in \Fig\ref{f1}c contributes to the classical correction to the
Minkowski metric
\begin{align}
\underline{\underline{g}}{}^{\cl}_{00}=(2+a)\,\frac{G^{2}m^{2}}{c^{4}r^{2}}\,.
\label{e23a}%
\end{align}
However, this correction is actually induced by the tree diagram
(\Fig\ref{f2}a). The decomposition on the background field and its fluctuations
(\ref{e2}) has no sense for such diagrams, because the integration momentum
(flowing through the "legs with crosses"\ in \Fig\ref{f2}a) is of the order of
$q$. It follows that the leading classical correction to the Minkowski metric

\begin{align}
\underline{g}{}^{\cl}_{00}=- 2\,\frac{Gm}{c^2 r}
\end{align}
is of the same order as the field $h_{\mu\nu}$; consequently, it serves no
purpose to distinguish them. Since the correction (\ref{e23a}) is not the
leading one, therefore it is possible to turn back to the initial variables
(\ref{e2}) rather than (\ref{e16}), i.\,e.
\begin{align}
\underline{\underline{g}}{}_{00}^{\hr}=\underline{\underline{g}}{}^{\cl}_{00}
-\frac{a}{4}\left(\underline{g}{}^{\cl}_{00}\right)^{2}=\frac{2G^{2}m^{2}}{c^{4}r^{2}}\,,
\label{e24}%
\end{align}

\noindent where $\underline{\underline{g}}{}^{\hr}_{00}$ is the second order
term in the expansion of the Schwarzschild metric in the harmonic coordinates.

It should be repeated once again that the quantum correction (\ref{e23}) is the
leading one, therefore the trick (\ref{e24}) does not permit to turn back to
the former variables, i.\,e, the correction must be invariant by itself. An
important point is that $a$-dependent contributions to the potential vanish in
the sum of the diagrams \Fig\ref{f2}a  and \Fig\ref{f2}b only, i.\,e., even on
the level of classical gravity one cannot introduce the physically meaningful
"one-particle-irreducible potential"\ (contrary to the section VIII of
\Ref\cite{Bjerrum-Bohr:2002ks}).

\begin{acknowledgments}
I would like to thank I.B. Khriplovich for his helpful comments and
discussions. The investigation was supported by the Russian Foundation for
Basic Research through Grant No. 05-02-16627-a.
\end{acknowledgments}


\end{document}